\documentclass[a4paper,fleqn]{cas-sc}

\usepackage[authoryear,longnamesfirst]{natbib}
\usepackage{multirow}
\usepackage{array}
\usepackage{rotating}
\usepackage[dvipsnames]{xcolor}
\usepackage{float}
\usepackage[most]{tcolorbox}
\usepackage[shortlabels]{enumitem}

\newcolumntype{L}[1]{>{\raggedright\let\newline\\\arraybackslash\hspace{0pt}}m{#1}}
\newcolumntype{C}[1]{>{\centering\let\newline\\\arraybackslash\hspace{0pt}}m{#1}}
\newcolumntype{R}[1]{>{\raggedleft\let\newline\\\arraybackslash\hspace{0pt}}m{#1}}

\def\tsc#1{\csdef{#1}{\textsc{\lowercase{#1}}\xspace}}
\tsc{PB}

\begin{document}
\let\WriteBookmarks\relax
\def\floatpagepagefraction{1}
\def\textpagefraction{.001}
\shorttitle{Numerical weather prediction at the crossroads}
\shortauthors{P. Bauer}

\title [mode = title]{What if? Numerical weather prediction at the crossroads}                      

\author[1]{Peter Bauer}[type=editor,
                        auid=000,bioid=1,
                        orcid=0000-0002-3205-6055]
\cormark[1]
\fnmark[1]
\ead{bauerspeter@icloud.com}

\credit{Conceptualization of this study, Writing - Original draft preparation}

\affiliation[1]{organization={Max-Planck-Institute for Meteorology},
                addressline={Bundessstrasse 53}, 
                city={Hamburg},
                postcode={20146}, 
                country={Germany}}




\cortext[cor1]{Corresponding author}


\nonumnote{This article presents scenarios for future numerical weather prediction operational computing through federated computing, data handling and machine learning.}

\begin{abstract}
This paper provides an outlook on the future of operational weather prediction given the recent evolution in science, computing and machine learning. In many parts, this evolution strongly deviates from the strategy operational centres have formulated only several years ago. New opportunities in digital technology have greatly accelerated progress, and the full integration of computational science in numerical weather prediction centres is common knowledge now. Within the last few years, a vast machine learning research community has emerged for creating new and tailor-made products, accelerating processing and - most of all - creating emulators for the entire production of global forecasts that outperform traditional systems at the spatial resolution of the training data. In this context, the role of both numerical models and observations is changing from being equation to data driven. Analyses and reanalyses are becoming the new currency for training machine learning, and operational centres are in a powerful position as they generate these datasets based on decades worth of experience. This environment creates incredible opportunities to progress much faster than in the past but also uncertainties about what the strategic implications on defining cost-effective and sustainable research and operations are, and how to achieve sufficient high-performance computing and data handling capacities. It will take individual national public services a while to understand what to focus on and how to coordinate their substantial investments in staff and infrastructure at institutional, national and international level. This paper addresses this new situation operational weather prediction finds itself in through formulating the most likely "what if?" scenarios for the near future and provides an outline for how weather centres could adapt. 
\end{abstract}


\begin{highlights}
\item The time-critical operational production of numerical weather forecast by individual centres is reaching affordable computational limits.
\item Machine learning promises alternative options but requires complementary investments in physics based simulations and reanalyses.
\item Future cost-effective analysis and forecast production needs a federated approach to extreme-scale computing and data handling.
\end{highlights}


\maketitle

\section{Introduction}\label{section:intro}
Numerical weather prediction (NWP) is the foundation for civil protection services to society, it serves many industries with dependencies on the environment, and it is in every citizen's mind when planning and managing their lives and well-being. With the growing impacts of extreme weather on lives and infrastructures in the context of climate change, the expectations on greater reliability of forecasts are increasing faster than ever before.

The steady progress of forecasting skill is well documented and known as the quiet revolution of NWP \citep{bauer2015quiet}. The meteorological community is likely one of the best organised science and service communities in terms of globally concerted investments in billion-dollar observing systems, the free and open exchange of data, the coordination of research and development, and the standardisation and quality control of outputs. These responsibilities are shared between national and international weather prediction centres under the umbrella of the World Meteorological Organization (WMO) \citep{brunet2023advancing}. The more recent thrust of private companies into this domain has further accelerated progress and led to a very diverse weather service ecosystem serving global and local needs at the same time. 

Digital technologies have always been one of the main enablers of NWP, mostly for performing the forecast model calculations fast enough to generate timely output. High-performance computing (HPC) has provided the backbone for forecast production since the 1950s. The key specifications of simulation models, e.g., spatial resolution, time stepping, Earth-system process complexity, and observational data volumes and diversity used in data assimilation have grown commensurate with the available (and affordable) computing power \citep{michalakes2020hpc} by a million times in the past 20 years. 

The almost natural progression of NWP skill following the evolution of computing technology is coming to an end, because we are reaching the upper limits of (i) processing speed for roughly the same cost according to Moore's law and Dennard scaling \citep{shalf2020future}, and of (ii) the achievable sustained computing performance of complex NWP codes on this technology. This means that further enhancements ultimately need bigger and more expensive machines. Their acquisition and operation cost are likely to become unaffordable for operational weather centres if forecast model specifications will grow at the same pace as in the past.

Machine learning (ML) promises alternative options if not a way out of this dilemma. The underlying methods have re--emerged after decades of dormancy because of a processor and software technology revolution that was stimulated by a vast range of commercial applications outside environmental science. The huge data amounts and computing capacities available today have turned 20th century's ML into 21st century's deep learning\footnote{For simplicity, we will use the term 'machine learning' (ML) in the remainder of the paper.}.

This situation creates an environment of great optimism but also fear of the public losing ownership over this important domain. It is fuelled by substantial public funding for ML and a huge push from private companies discovering the profits to be made in weather and climate services. ML offers accelerated progress to the public players but requires new levels of science and technology coordination, including the need to respond faster to what technology has to offer, and how research can be translated into operational benefit while still maintaining the community's quality standards \citep{frolov2024integration,bauer2023deep}. 

While it is easy to see {\it why} weather centres will need to adjust quickly and, in parts, radically to the rapid developments, it is more difficult to see {\it how} operational predictions will actually be done in practice in the future. For this purpose, this paper describes what-if scenarios and possible pathways towards them:

\begin{itemize}
    \item What if the quiet revolution has reached its limits and large km-scale ensemble forecasts -- the scientific reference target -- cannot be run operationally on the next two generations of supercomputers at weather centres?
    \item What if both data assimilation used for creating initial conditions and forecasts based on ML will be superior to physics-based methods for operational predictions?
    \item What if the present engagement of 'big tech' companies with NWP will continue and result in alternative commercial operational products?
    \item What if ML models continue to grow in size (e.g., via the use of foundation models) and the training of ML applications will become the largest HPC application in Earth science?
\end{itemize}

For the rest of the paper, we assume that all of the what-if scenarios become reality to address our headline concern, namely that computing and data handling cost will become unaffordable following traditional ways. If adjusted, the computing cost of the actual operational forecasts would be reduced significantly when using ML because the computing cost of ML inference -- the application of trained ML algorithms to new input -- is negligible. The main drivers for compute power for NWP would be the generation of reference training datasets -- high-resolution model simulations and reanalyses -- and ML training. As these tasks will be too big for the local supercomputing facilities at individual weather centres, the following adaptations by the NWP community in Europe and world-wide would become necessary:

\begin{enumerate}[(A)]
    \item The time-critical production suite for both initial conditions and operational forecasts will be based on ML inference.
    \item Reference dataset production cycles will be created with frequent updates to generate the next-generation training datasets.
    \item The costly generation of training datasets will be shared between operational centres and third-party programmes to optimise the use of national and international computing resources and to democratise the outcomes.
    \item The provision of both intellectual and digital infrastructure resources, and quality-controlled data, will be governed through a sustainable public-private partnership framework.
    \item Software and data management capability and challenges will be addressed community wide, and data handling will be addressed as an internationally federated effort.
\end{enumerate}

In Section \ref{section:why}, the paper will provide arguments why the what-if scenarios are realistic and explain in more detail what the anticipated changes will mean for the NWP community in Section \ref{section:how}.

\section{Why the what-if scenarios are likely}\label{section:why}
\subsection{Background on forecast skill}
Today, operational centres tend to produce longer-term, visionary strategies with a 5-10-year focus, and then more detailed, technical implementation plans including targets for operational system version upgrades for 1-5 years. The former align with regular core budget cycles and big investments such as new HPC infrastructures while the latter align with institutional and (inter)national research and service programmes. 

Strategies are guided by service commitments that advance the value for society returned from the investments made in public institutions, and the definition of science ambitions and technology requirements to achieve the expected service level. The national centre strategies also account for the role of the international service ecosystem and commercial providers, and include the research and development funding opportunities offered from third-party programmes. Here, WMO plays an important role for international strategic research collaboration and for the coordination of long-term commitments to Earth observation, but also for promoting the outcomes in more and less developed countries.

\begin{figure}[h]
    \includegraphics[scale=0.55]{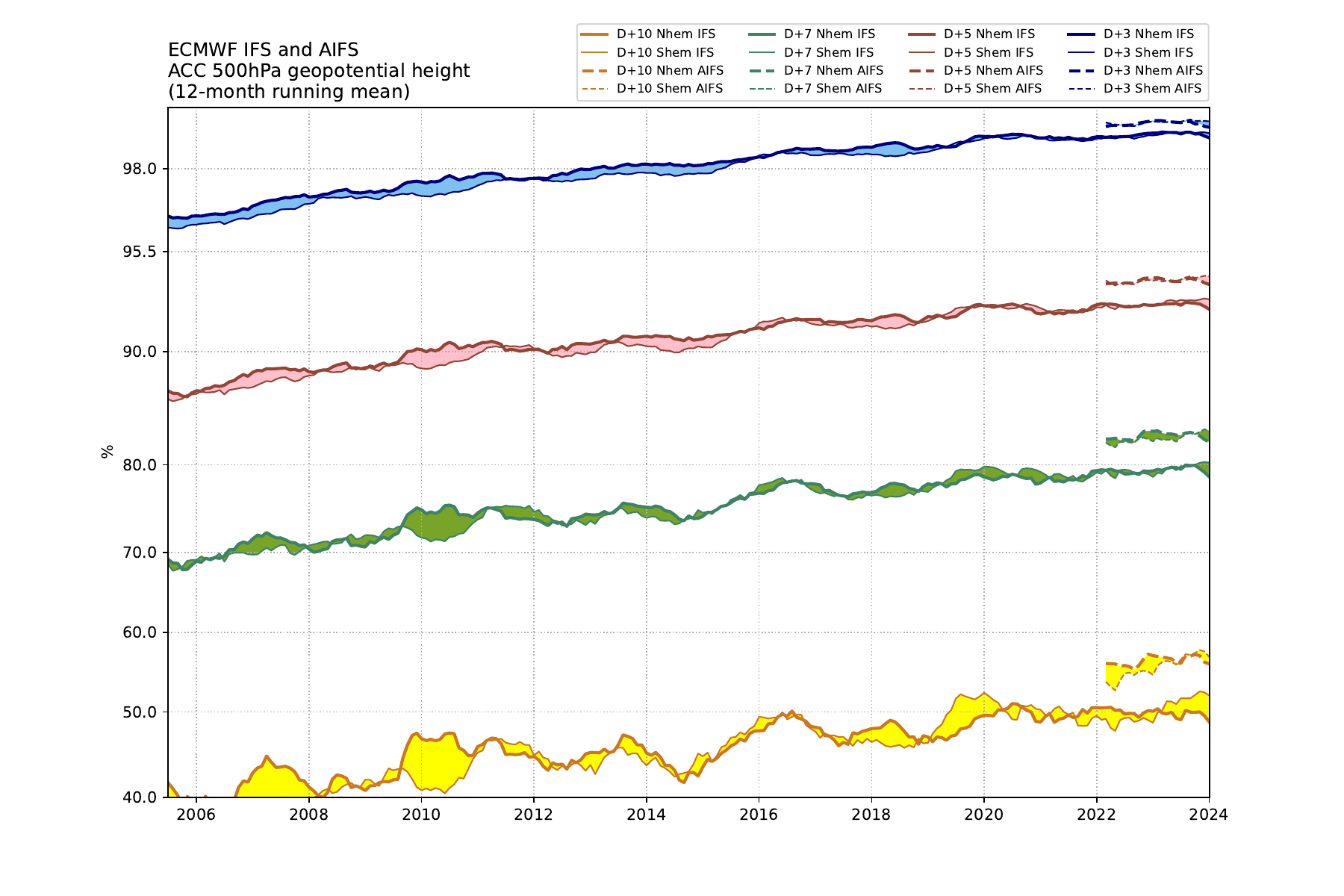}
    \caption{Time series of 12-month running mean forecast skill of the ECMWF high-resolution deterministic Integrated Forecasting System (IFS, solid lines) and ML IFS (AIFS, dashed lines). Skill is expressed as the anomaly correlation (= correlation between forecasts and verifying analyses, normalised by climatological signal) for the 500 hPa height. Thick (thin) curves show northern (southern) hemisphere averages for day-3 (blue), day-5 (red), day-7 (green) and day-10 (yellow). (Figure courtesy Martin Janousek, ECMWF)}
    \label{fig:scores}
\end{figure}

Today's shorter-term implementation planning is based on the assumption that science and technology developments need to advance at a steady pace, that the core of the operational production is based on numerical systems to be run over time-critical schedules on big machines preferably owned by NWP centres, and that weather and climate derivative services inherit outcomes and access to infrastructure to also help produce their next versions of service suites. 

Commercial company business models mostly create added value on top of public data output, but there are exceptions where large companies aim to run independent prediction systems (like The Weather Company) and act as primary Earth observation data providers (like SPIRE). This approach has paid off in the past and established a framework of predictable progress and a well-funded and acknowledged public service space complemented by sufficient opportunities for business. 

Weather forecast skill is measured by comparing forecasts with observations and analyses. The latter are produced by combining short-range forecasts with observations through data assimilation. Observations are sparse in space and time and constrain a physics based, dynamical model to describe the physically consistent evolution of all weather parameters at all grid points. These analyses are also used to initialise the next forecasts. Reanalyses apply this process to decades worth of observations using the same numerical model and data assimilation system to provide climate monitoring (e.g. \citet{hersbach2020era5}).

The continual cycle of observing, analysing and forecasting feeds forecast verification with solid statistical data from days to seasons. Figure~\ref{fig:scores} shows a well-known example of NWP forecast skill evolution from the European Centre for Medium-Range Weather Forecasts (ECMWF) Integrated Forecasting System (IFS) and its ML counterpart (AIFS, \citet{aifs}). The graphs displays the synoptic-scale 500 hPa geopotential height in the northern and southern hemispheres, which represents a metric to denote the skill to predict weather-patterns such as fronts. All operational centres produce such skill monitoring and contribute to well established global performance assessments using the same standards. 

The figure illustrates the steady growth of medium-range skill, here at days 3/5/7/10, as reported in past publications \citep{bauer2015quiet}. The ML-based AIFS (blue) has outperformed the IFS for this specific skill score straight away since its publication in 2022 \citep{aifs}. Note that these skill score evaluations are produced at coarser than the full model resolution.

The cornerstones of skill improvement are well documented: stable and accurate numerical schemes to solve the equations of motion and thermodynamics, so-called physical parametrisations that describe the impact of subgrid-scale processes like radiation and clouds on the resolved scales, data assimilation methods combining simulated and observed data to produce best estimates of the state of the system at a given time, and ensemble methods to characterise uncertainties and how they propagate from initial states through forecasts. Increasingly, weather models have included land, sea-surface wave and ocean sub-models as these are relevant for weather, and more seamlessly represent processes ranging from days to seasons. The AIFS benefits from turning decades worth of reanalyses data into enhanced skill when initialised with the same daily initial conditions as the IFS.

Steady improvements in the past meant that centres, supported by the wider research community, aim to improve these components within the existing framework and without radically changing the underlying numerical systems. Improvements in the observations as well as the numerical methods and algorithms that are used for both the forecast model and data assimilation have led to significant skill over time, but a leading driver for improvements is the increase of spatial resolution as it reduces the range of unresolved and therefore approximated processes with immediate benefits when describing flow in mountainous and coastal areas, deep convection and cloud formation, but also atmosphere-land interaction and waves \citep{wedi2014increasing}. The km-scale, limited-area prediction systems that all national centres maintain to produce more reliable local forecasts are testimonies to this fact. 

At present, the ultimate goal for global prediction is to resolve storms and even cloud systems as these are responsible for the vertical energy transfer and interact with the three-dimensional large-scale circulation influencing weather globally. This goal translates to spatial resolutions of 1 km or better \citep{stevens2019dyamond,wedi2020baseline,hohenegger2023icon}, and is also relevant for climate simulations predicting the future of global weather \citep{palmer2019scientific,hewitt2022small,rackow2022delayed}. Such simulations are targeted in programmes such as Destination Earth in Europe \citep{DestinE}. There is also an opportunity to bridge to the high-resolution modelling efforts in preparation for the next Coupled Model Intercomparison Project (CMIP-7) \citep{roberts2024high}, and unify simulation protocols and evaluation standards.

Global km-scale forecasts also require global km-scale data assimilation to create the initial conditions. This is presently not only impossible due to HPC limitations but also because today's data assimilation methods are not equipped for dealing with non-linear processes acting across this range of scales \citep{carrassi2018data}. The same data assimilation methods are also used to create reanalyses, for which the same considerations apply. Lastly, the km-scale vision  also includes the need for quantifying uncertainty through ensembles \citep{palmer2020vision}, which inflates the computational load even more. 

ECMWF is, for example, currently running 51-member ensemble forecasts at 9 km resolution operationally several times per day. If their compute budget would remain at a similar level as today, the expected improvements in computing power from the next generations of supercomputers would not suffice to push the resolution of operational ensembles to 1 km. This is because a km-scale ensemble would approximately be $>$100 times the current cost given the increase of the number of grid points and smaller time steps required to capture smaller-scale and faster processes. This makes km-scale storm resolving and even finer cloud resolving simulations result in nearly insurmountable computational hurdles for operational global weather and climate prediction alike \citep{schulthess2018reflecting}.

\subsection{Background on high-performance computing}
\subsubsection{Modus operandi}
The traditional approach to the acquisition and operation of HPC resources of most national centres in Europe, USA, Japan, China, India, but also ECMWF is to buy or lease a system over which the centres have full control during the contracted lifetime. There are deviations from this principle where several, smaller centres share a system (e.g., Nordic countries in Europe) or manage an allocation on a larger system that is owned by another public entity (e.g., Bureau of Meteorology in Australia, MeteoSwiss in Europe). These HPC systems usually comprise both computing and data handling resources and are contracted for four years or longer with HPC providers. 

These HPC procurements have a lead-time of at least two years used for exploring the target requirements and potential vendors, and about one year is spent on system installation and acceptance testing before the active operational period starts. The most important disadvantage of this approach is that technology moves fast and that the HPC system contains already outdated components by the time it becomes operational. While selected components can be upgraded in phases during contracts, the basic architecture remains unchanged, and vast deviations from a chosen computing/data handling configuration with implications on networks and power/cooling supply are impossible. The growing complexity of both NWP codes and HPC systems makes the cost-effective transfer of new science into operations therefore increasingly difficult.

More recently, NWP centres have started to outsource some of their post-processing and data handling to commercial providers using cloud-based services. Some of these use a software service platform owned by selected NWP centres serving the wider community (e.g. European Weather Cloud). The Copernicus services in Europe also piggy-back on the operational centre systems by paying for computing and data handling allocations, and by contracting their own cloud-based services in their vicinity. This makes the adaptation of codes based on NWP software easier and avoids costly data transfers to separate infrastructures.

The first operational centres to break away from the prime HPC-ownership model have been NOAA and the UK Met Office. NOAA sub-contracted General Dynamics Information Technology (GDIT) \citep{NOAAGDIT} in 2020 and the Met Office sub-contracted Microsoft in 2021 \citep{MetOfficeMicrosoft}. Both are ten-year commitments with total envelopes over \$500 million and £1.2 billion, respectively. The expectation is that HPC-as-a-service will reduce the centre-internal spin-up and enhance technological agility, eventually delivering better value for the investment. In how far this choice is economical and independent from vendor specific solution lock-in remains to be seen, but HPC-as-a-service appears to be scalable beyond what is presently available through the institutional ownership model.

\begin{table}[!h]
\centering
    \caption{Key data of selected HPC systems from operational NWP and other centres (ECMWF: European Centre for Medium-Range Weather Forecasts, NOAA: National Oceanic and Atmospheric Administration, JMA: Japan Meteorological Agency, KMA: Korean Meteorological Agency, CSC: Finnish Information Technology Center for Science (LUMI: Large Unified Modern Infrastructure), CINECA: Consorzio Interuniversitario del Nord-Est per il Calcolo Automatico, BSC: Barcelona Supercomputing Center, CSCS: Swiss National Supercomputing Centre, DOE: Department of Energy). $^1$Note that performance figures may denote first phase of a system to be upgraded in future phases of a contract.$^2$Performance achieved with High-Performance Linpack (HPL) benchmark in double precision \citep{dongarra2003linpack}.}     

    \begin{small}
    \resizebox{0.9\columnwidth}{!}{\begin{tabular}{|l|l|l|l|}
    \hline
    {\bfseries Centre, country} & {\bfseries Sustained performance$^{1,2}$} &{\bfseries Power consumption}  & {\bfseries Main vendor/chip provider:}\\
    & {\bfseries (Pflop/s)} &{\bfseries (MW)}  & {\bfseries (Nodes)}\\
    \hline
    ECMWF, Europe                    &   4x7 & 4x1.2 & Atos/AMD: 4 x 1,920 CPU \\
    Met Office, UK                   &    50 & N/A  & HPE/AMD: N/A    \\
    NOAA, USA                        &   2x10 &  N/A  & HPE/AMD: 2 x 2,560 CPU \\
    JMA, Japan                       & 2x13.4 & 2x0.9 & Fujitsu/Fujitsu: 9,216 CPU    \\
    KMA, Korea                       &   2x18 & 2x3.3 & Lenovo/Intel: 2 x 4,023 CPU  \\
    \hline
    CSC, Finland (LUMI)       &   380  &   7.1 & HPE/AMD: 2,048 CPU, 2,978 GPU)  \\
    CINECA, Italy (Leonardo)  &   241  &   7.5 & Atos/NVIDIA: 1,536 CPU, 3,456 GPU \\
    BSC, Spain (MareNostrum5) &   175  &   4.2 & Atos/Intel\&NVIDIA: 6,408 CPU, 1,120 GPU  \\
    CSCS, Switzerland (Alps)         &   270  &   5.2 & HPE/AMD\&NVIDIA: 1,024 CPU, 2,688 GPU \\
    Riken, Japan (Fugaku)            &   442  &  29.9 & Fujitsu/Fujitsu: 158,976 CPU \\   
    DOE, USA (Frontier)              & 1,206  &  22.8 & HPE/AMD: 9,408 GPU \\
    \hline
       \end{tabular}}
    \end{small} 
\label{tab:HPC}
\end{table}

Table~\ref{tab:HPC} summarises the key computing figures of the largest NWP centres and adds, for comparison, the leading European HPC systems co-funded by the European Commission through the EuroHPC programme, the Department of Energy's (DOE) Frontier and the Japanese Riken's Fugaku machines. The leading general-purpose systems provide between 150-300 Pflop/s (10$^{15}$ (Peta) floating point operations per second at 64 bit precision) sustained performance, a performance which is mostly derived from benchmarking idealised numerical problems on large processor allocations that maximise computational intensity \citep{dongarra2003linpack}. 

Most of the present operational NWP suite and workflow set-ups already stretch the existing system capacities. First, the prediction suites produce the initial conditions (called analyses) for forecasts based on the latest incoming observations and short-range forecasts from previous cycles. This is done multiple times per day. The computational load for analyses and medium-range forecasts is very similar as the former are based on solving a complex global four-dimensional optimisation problem ingesting 100s of millions of new observations, and the latter solves the complex numerical Earth-system equation framework in thousands of time steps for the next days to weeks. Both analyses and forecasts also include separate, ensemble based suites that produce uncertainty estimates, and these dominate the computing cost as they multiply the cost by as many ensemble members are used.

This means that the analyses and forecast suite configurations determine the required number of compute nodes for each analysis and forecast cycle. As these need to be executed as quickly as possible, the number of allocated compute nodes is maximized. This allocation needs to fit into the maximum capability of the available HPC system, that is the maximum available node allocation. The integration of all suites over any given day defines the required capacity (in node-hours).

An example: in operations ECMWF runs twice per day a 51-member data assimilation ensemble (EDA), a 51-member ensemble medium-range forecast (ENS) for days 1-15, complemented by another ensemble forecast generating the limited-area model boundary conditions (BC) for days 1-6. In addition, there is one monthly forecast for days 1-46 per day and two sets of reforecasts (REF) per week that are used to calibrate for systematic errors in the ensembles \citep{Cycle49r1}. 

At 9 km spatial resolution, each member of ENS uses about 50 compute nodes (plus a few nodes for input and output) \citep{Cycle49r1HPC}. Completing the forecasts in about one hour requires therefore at least 2,550 nodes. The on-the-fly product-generation runs in parallel and occupies at least 100 nodes \citep{NextGenIO}.

As the EDA runs its outer-loop trajectories at the same resolution as ENS, the total node allocation is about the same as for ENS. This also applies to the BC suite, but it requires less time to complete due to the shorter forecast range. The reforecasts require a smaller allocation because they can use more time to complete. However, their capacity requirement is substantial as each execution with 11 members accumulates statistics over 20 years, and thus consumes over 11,000 node-hours of capacity.

The two ECMWF HPC clusters available for operations comprise a total of 3,840 nodes (see Table~\ref{tab:HPC}). Leaving several hundred nodes free for emergency measures on each cluster means that the allocations for EDA, ENS and BC already claim 80-85\% of the available capability.

\subsubsection{The affordability limit}
Projecting the capacity into the future is revealing the challenge. The procurement of the next-generation HPC systems is usually based on a 4--5-year extrapolation of the analysis and forecast suites in terms of spatial resolution, complexity and ensemble size. At ECMWF, this has led to a doubling of horizontal resolution of the single, deterministic forecast in about 24 months in the past \citep{schulthess2018reflecting}. However, if this rate would be maintained, and the number of ensemble members would stay the same, the next generation of HPC systems at national centres would approach the size of the present EuroHPC infrastructures (see Table~\ref{tab:HPC}), requiring acquisition budgets in the range \texteuro 150-250 million. The requirements for data handling go hand-in-hand with those for computing as the daily operational output will reach 1 PByte very soon. 

It is not obvious that publicly funded centre budgets will be able to afford such amounts given the economic conditions in times where other geo-political pressures dominate. As a baseline, this means that individual NWP centres may struggle to afford HPC systems that can deliver more than a 100 Pflop/s (10$^{15}$ floating-point operations per second) and that consume more than 10 MW. For the first time, HPC capability and capacity will reach a hard affordability limit.

On a general principle, HPC limitations to progress in NWP are not new. Both weather and climate communities have invested in augmenting the computing performance of their simulations for decades. The concerns about the scalability of complex Earth-system models, particularly on heterogeneous processor and memory technologies, has produced several joint projects at both national and international level \citep{lawrence2018crossing,schulthess2018reflecting,muller2019escape,bauer2021digital,govett2024exascale}. 

Realistically though, present-day models produce only a few percent sustained floating-point performance because the ratio between computations and data movements is fairly poor \citep{carman2017position}. Investments in parallelisation and memory access improve this ratio, and a factor of 10 and more acceleration gains can be obtained for individual model components but not necessarily for entire models \citep{muller2019escape,dahm2023pace}. These efforts are only partly supported by generic programming model options though and involve deep-dives into algorithms and code design. Porting model components to other, more specialised processors like field programmable gate arrays (FPGA) are possible and may produce good performance and energy efficiency gains. However, they only work for specific tasks, are tedious to carry out and almost impossible to generalise \citep{targett2021systematically}. 

While the community has spent significant efforts and made very good progress to make global NWP models portable and more efficient, the step-change that meets the above stated requirements cannot be expected only from code adaptation to heterogeneous processor and memory technologies in the coming years. This translates into larger systems and more cost.

An important aspect is electrical power consumption that ranges from about $\approx$5 MW for the operational centre systems to $\approx$30 MW for the leading flagship computers shown in Table~\ref{tab:HPC}. This translates to about \$5-30 million energy cost per year, ideally to be drawn from renewable resources. This can be more easily achieved in northern countries where free cooling is more effective and wind and hydro-power sources are abundant. But this is not the case for the existing systems. As the energy-per-flop ratio is not decreasing for modern HPC hardware and more flop/s are needed for running higher-resolution and more complex models, the overall power consumption at most compute sites will also approach affordability limits. Apart from affordability, the need for providing environmentally sustainable solutions makes it likely that weather centres will have to move their HPC facilities to colder regions with sufficient wind, solar and hydropower provisions in the future.

\subsection{Background on machine learning}
\subsubsection{Recent developments}
In recent years, computing technology supporting commercial artificial intelligence (AI) applications has boomed. Easy-to-use and efficient software environments became available, and large public funding programs have appeared. This environment has stimulated a large wave of research projects. It is also fuelled by substantial commercial investments that spill into NWP. Consequently, the data-based training for solving larger and more complex tasks has grown at an unprecedented rate. 

During the initial phase of this development, weather and climate centres focused on hybrid ML applications that couple ML tools with the conventional prediction workflow, while big technology companies, including NVIDIA, Google Deep Mind, Huawei and Microsoft were the first to realise that ML models that emulate the entire forecast production can actually compete and, in fact, outperform physics-based NWP systems for both deterministic and ensemble prediction \citep{fourcastnet,pangu,graphcast,gencast}. The capability to build such models was clearly underestimated by the weather community when initial feasibility experiments were performed \citep{dueben2018challenges}. 

Consequently, ML has changed the NWP landscape substantially during the last two years, and has had an impact on the basic understanding of how forecasts are produced. Weather and climate modelling centres are now catching up quickly -- for example, ECMWF has pushed their first global ML model (called AIFS) into semi-operational use \citep{aifs}. The quality of predictions in terms of forecast scores of the AIFS represents a step-change when compared to physics based models (see Figure~\ref{fig:scores}) and pure ML models are able to predict many kinds of extreme events and show much more physical consistency than originally expected by domain scientists \citep{ZiedPangu}. 

\subsubsection{Today's frontier}
It is unlikely and undesirable that the use of physics based NWP models stops. They provide the physical-process based insight that is necessary to understand skill, which makes their continuous development a high priority. They are also needed to represent singular events, for example aerosol injections into the atmosphere from volcanic outbreaks, that ML models are more difficult to train for. 

However, the negligible cost of running ML models seem to mark the end of the quiet revolution of physics based models in terms of daily, time-critical production. Today's business-as-usual approach will be replaced by a new generation of ML models that produce most of the daily predictions for end-users in the near future. The result will be a two-tiered approach where ML models perform the urgent and specialised tasks, and the numerical models produce scientific insight and reference data.

One of today's most important questions is whether ML can also be used for data assimilation as successfully as for forecast modelling. In terms of information content, it seems hard to imagine that the presently available observations, despite comprising 100s of millions of satellite and conventional data points per day, can describe the four-dimensional state-space of a global Earth-system at km-scale resolution and hundreds of vertical levels. However, ML appears to be suitable to replace many of the individual steps that are performed in data assimilation such as observation operators, observation quality control, error estimation for observations and the model, the interpolation in space and time, and the blending of information from observations and the model \citep{geer2021learning}. ML also promises to improve on the exploitation of observational information as today's analysis systems only exploit 5-10\% of the truly available observational data volume (e.g. \citet{ECMWFmonitoring}). It is therefore not a big surprise that first tests where ML performs the full data assimilation process provide very promising results \citep{huang2024diffda,aardvark,sun2024fuxi}.

The next challenges for ML in the weather prediction domain are the push of ML models to represent the full Earth system including land, ocean, sea-ice, and waves. First ML-ocean models are already available \citep{xihe}. This also leads to a closer investigation of the usability of ML models for climate predictions noting that first, so-called Atmospheric Model Intercomparison Project-type \citep{gates1999overview} simulations have already been successful \citep{neuralgcm}. Another question is whether hybrid approaches that couple a ML model with a conventional model \citep{neuralgcm} will eventually be outperformed by pure ML methods in terms of forecast scores and forecast consistency. ML methods enhancing conventional-model output also exist, for example, to perform online bias corrections within the model simulations \citep{Bonavita,Laloyaux}, or for post-processing and spatial down-scaling \citep{harris2022,bouallegue2024}.

Currently, the cost for the training of pure ML models is still substantially lower than executing high-resolution physics based models. For example, training the NeuralGCM ensemble output at 0.7 degree spatial resolution once over 3 weeks on 256 Tensor Processing Units (TPU) \citep{neuralgcm} is much less computationally intensive in terms of processor-hours than the above stated allocation of 2,550 nodes needed for the daily ensemble production at ECMWF. However, as the ML models are pushed to higher resolution and are trained for ensemble forecasts using, for example, diffusion and ensemble score-based loss functions \citep{gencast, pacchiardi2024}, the cost for training is increasing rapidly. 

Furthermore, the trend in ML learning is moving towards so-called foundation models. These are ML models that can be used for several application areas and are trained using several different input and output data types in a representation learning approach. Their training is gap-filling the dataset and therefore learning to probabilistically transfer different input streams in space and time to retrieve the chosen outputs. The first results with such foundation models in Earth system science are also very promising \citep{atmorep,climax,aurora}. 

As the training datasets and networks for foundation models are much larger when compared to task-based models, they have the potential to become the largest HPC application in the domain of weather and climate predictions in the near future \citep{orbit}. Here, the mostly 16-bit TPU based training of foundation models is different from the mostly 64-bit CPU-GPU based numerical-model-execution computing challenge. Foundation models are likely to have a significant impact on how ML is used for weather forecasting, but they still need to show that they can eventually produce better results for important task-based applications when compared to pure ML-emulator NWP models. 

\subsubsection{Public-private collaboration}
An important aspect of ML research is that it has already established a wide public-private collaboration environment where private actors aim to create and exploit new business opportunities that were not accessible to them in the classical NWP world. Public actors, of course, understand the huge potential for accelerating progress and benefit from substantial Artificial Intelligence (AI) funding programmes that national governments release in order to remain competitive. The work and power distribution between public and private entities has therefore changed \citep{bauer2023deep}, in particular, as some of the big technology companies seem determined to turn their research results into payable operational weather products for their users. But if commercial companies stop sharing their data-driven models and start patenting their products \citep{cheon2023climate,ClimateAI}, this public-private collaboration environment may change yet again. 

One specific aspect of this collaboration will be the added capabilities for user interaction and the interrogation and interpretation of data through large-language models (LLM). This adds new levels of complexity to model design, training data and cost, but also to data and application governance \citep{bauer2023deep}. There are also important ethical considerations for protecting the source and quality of data and outputs \citep{mcgovern2022we}. However, this topic is beyond the scope of this paper. 

\section{How would NWP centres adapt?}\label{section:how}
Given today's situation, the main question therefore is: how can operational NWP centres create an affordable environment to maintain and even accelerate progress realising the urgent need to deliver better services for a society that is increasingly exposed to the impacts of extreme weather? 

Help could come from an alternative scenario to the present operational practice towards a more efficient and agile technical prediction system set-up that we have already introduced in Section~\ref{section:intro}. The scenario would also boost scientific research, reduce the pressure for managing operational workloads at NWP centres, and create a more cost-effective use of digital technologies and HPC resources across institutional and national borders:

\begin{enumerate}[(A)]
    \item The time-critical production suite for both initial conditions and operational forecasts will be based on ML inference:
    \begin{itemize}
        \item the inference suites would comprise global, regional and local scales, and would include suites targeted at, for example, hydrology, air quality, agriculture or coastal management, but also boundary conditions for limited-area systems;
        \item the revision cycles would be shortened for faster uptake of new research and data, and include on-demand suites that are only activated when necessary, for example in situations leading towards a flooding event;
        \item the inference suites (and the necessary input datasets) would be open source and easily transferable between centres, HPC systems, and countries;
    \end{itemize}
    \item Reference dataset production cycles will be created with frequent updates to generate the next-generation training datasets:
    \begin{itemize}
        \item several advanced reference data generation suites including reanalyses, advanced model simulations and observational data records would be produced in parallel depending on the training target - global vs local, weather vs climate;
        \item a specific effort would be created to identify, promote and accelerate the development of advanced numerical models and data assimilation methods that offer the best physical realism and perform well on advanced HPC systems;
        \item beyond the consolidation of the already available and openly accessible observational framework, a specific effort would be made for ingesting experimental, internet-of-things (IoT), and commercial datasets as fast as possible.
    \end{itemize}
    \item The costly generation of training datasets will be shared between operational centres and third-party programmes to optimise the use of national and international computing resources and to democratise the outcomes:
    \begin{itemize}
        \item the allocation of resources at individual centres would remain reasonable while the collective allocation would be commensurate with the required capabilities/capacities;
        \item sustainable funding programmes external to NWP centres would be included, for example national/international research and development funds, space agencies, and digital-technology programmes;
        \item digital technology companies would contribute to the swift adaptation of workloads to new technologies and provide access to large-scale prototype HPC facilities;
        \item there would be dedicated resources to access the large-scale datasets for the production of user specific and on-demand ML suites.
    \end{itemize}
    \item The provision of both intellectual and digital infrastructure resources, and quality-controlled data, will be governed through a sustainable public-private partnership framework:
    \begin{itemize}
        \item the global public space would be shared by a few high-performance generic dataset producers and many specialised data producers/users that also feed user specific, tailormade suites for a wide range of local and topical applications into the community;
        \item extreme-scale HPC resources would be drawn from selected national and international centres that also serve other scientific communities, but where weather (and climate) data production obtains sufficiently large and sustainable allocations;
        \item the input data and output product quality definition would build on existing community standards to fulfil trustability expectations; 
        \item private companies would support the public parties with software services but also provide anonymised observational and application specific data; in turn, private companies would have access to quality controlled generic data generated by public entities.
    \end{itemize}
     \item Software and data management capability and challenges will be addressed community wide, and data handling will be addressed as an internationally federated effort:
     \begin{itemize}
         \item the output data management would be handled through a decentralised infrastructure approach with large active data spaces near the main production sites for fast data analytics access;
         \item there would be dedicated high-speed network links between selected high-performance data producers to facilitate access to large holdings;
         \item data-intensive ML based workloads would be run on/near the large active data spaces;
         \item federated compute power would be available next to the federated datasets and post-processing would include cloud based services;
         \item reference datasets (reanalyses, model simulations, observations) could feed into foundation models; all data would employ the same API and uncertainty representation methodology;
        \item weather and climate domain specific text and knowledge repositories would be created, maintained and enhanced that feed community LLMs helping to interpret prediction system output and to turn data into information.
     \end{itemize}
\end{enumerate}

Such a scenario clearly deviates from the present philosophy of only relying on a single, entirely physics-numerical methods based prediction system that generates analyses and forecasts at all scales and serves all downstream applications on time-critical schedules -- all on the same, local HPC infrastructure. The scenario does not abolish such a system but rather introduces a clear distinction between what needs to run in near real-time, how ML training data is generated and how the remaining numerical-method based systems are managed. It is unlikely that physics based predictions will be removed entirely from operations in the near future, but the fraction of compute power used for time-critical numerical vs ML models will shift significantly.

Our scenario also introduces the notion of a much wider reaching collaboration framework that is necessary to share workloads and make the best use of the collectively available HPC resources rather than the single centre--single production--single HPC infrastructure thinking. The need for sharing the multi-lateral computing resources also implies a more democratised approach to data and software for users everywhere. Eventually, this will also allow a much easier transfer of knowledge and value to less developed countries.

Table~\ref{tab:Outlook} shows the proposal for a timeline of the evolution from present-day operations towards our scenario. The node allocations are ballpark figures based on the existing performance and what could be expected from the acceleration by ML. They are only intended to serve as a guideline to the orders of magnitude needed to produce the different conventional model and assimilation as well as ML training tasks.

The main message of this table is that operational centres could soon achieve smaller node-hour allocations for their time-critical tasks while still investing in both numerical systems and the upgrades to ML-based suites. The heavy workloads requiring larger machines than affordable for individual NWP centres will increasingly be moved to external HPC centres, and duplication between centres be eliminated. These are storm-resolving ensemble simulations at km-scale (called $x$km in the table) and reanalyses. The daily processing of observations mostly refers to space and meteorological agencies while the regular production outside NWP centres refers to Tier 1 \& 0 HPC infrastructures maintained by large national/international programmes.

The timeline in Table~\ref{tab:Outlook} is certainly ambitious but it anticipates, first, the pace at which ML algorithm development is progressing and should be turned into operational benefits and, second, the imminent need to adjust HPC infrastructure procurement specifications at both national and international level. Demonstrating such a transition for selected centres to begin with would already be a great success.

\begin{sidewaystable}[htbp]
\caption{Scenario for three stages of future evolution of NWP-centre operational production in collaboration with observational data providers and HPC centres. Research experimentation and research-to-operations testing is not included. Node allocations are rough estimates based on experience with today's technology. Numbers in brackets denote duration of calculations. GPU nodes include CPU processors. $x$km refers to resolutions at km-scale.}     

\resizebox{\columnwidth}{!}{\begin{tabular}{ |L{2.5cm}|L{5cm}|L{5cm}||L{5cm}|L{5cm}|  }

 \hline
 & \multicolumn{2}{|c||}{{\bf Operational NWP centres}} & 
 \multicolumn{1}{c|}{{\bf Observational data providers}} & \multicolumn{1}{c|}{{\bf HPC centres}} \\
 & {\it Time-critical daily cycles (duration)} & {\it Regular production (duration)} & {\it Daily cycles (duration)} & {\it Regular production (duration)} \\
 \hline
 Today & \begin{itemize}[leftmargin=*]
         \item Observation processing, numerical analyses \& forecasts: 1,000-5,000 CPU nodes (hours)
         \item Experimental ML forecast inference: 1 GPU node (minutes)
         \end{itemize} 
       & \begin{itemize}[leftmargin=*]
         \item Numerical reanalyses: 500 CPU nodes (years)
         \item Experimental ML forecast training: 10s GPU nodes (weeks)
         \end{itemize} 
       & \begin{itemize}[leftmargin=*]
         \item Observational data gathering and pre-processing \& parameter retrievals: 100 CPU nodes (hours)
         \end{itemize} 
       & \begin{itemize}[leftmargin=*]
         \item Observational data reprocessing: 100 CPU nodes (months)
         \item Numerical $x$km-scale simulations: 1,000 GPU nodes (months)
         \end{itemize} \\
 \hline
 \hline 
 Shorter-term \newline (1-3 years) 
       & \begin{itemize}[leftmargin=*]
         \item Observation processing, ML accelerated numerical analyses, ML forecast inference: 1,000 CPU-GPU nodes (minutes-hours)
         \item Experimental ML analyses inference: 10 GPU nodes (minutes)
         \item Experimental ML $x$km-scale forecast inference: 10 GPU nodes (minutes)
         \end{itemize} 
       & \begin{itemize}[leftmargin=*]
         \item Numerical reanalyses: 500-1,000 CPU nodes (years)
         \item Experimental ML analysis training: 100s GPU nodes (weeks)
         \item Experimental ML $x$km-scale forecast training: 100-1000 GPU nodes (weeks)
         \end{itemize} 
       & \begin{itemize}[leftmargin=*]
         \item Observational data gathering and pre-processing, ML parameter retrievals: 100 CPU nodes, 10 GPU nodes (minutes-hours)
         \end{itemize} 
       & \begin{itemize}[leftmargin=*]
         \item Observational data reprocessing: 100 CPU nodes (months)
         \item Experimental ML reanalysis training and inference: 100-1000 GPU nodes (months)
         \item ML accelerated numerical $x$km-scale simulations: 1,000-10,000 GPU nodes (months)
         \item Experimental ML foundation models: 100 GPU nodes (weeks-months)
         \end{itemize} \\
 \hline
 \hline 
 Longer-term \newline (3-5 years) 
       & \begin{itemize}[leftmargin=*]
         \item Observation processing, ML analyses \& $x$km-scale forecasts: 10-100 GPU nodes 
         \end{itemize} 
       & \begin{itemize}[leftmargin=*]
         \item ML accelerated numerical reanalyses: 500-1,000 CPU nodes (months)
         \item ML reanalysis inference: 100-500 GPU nodes (months)
         \item ML analysis and reanalysis training: 100 GPU nodes (weeks-months)
         \end{itemize}
       & \begin{itemize}[leftmargin=*]
         \item Observational data gathering and pre-processing, ML parameter retrievals: 100 CPU nodes, 10-100 GPU nodes (minutes-hours)
         \end{itemize} 
       & \begin{itemize}[leftmargin=*]
         \item Observational data reprocessing: 100 CPU nodes
         \item ML $x$km-scale simulations: 1,000 GPU nodes (months)
         \item ML foundation models: 1,000 GPU nodes (months)
         \end{itemize} \\
 \hline
\end{tabular}}
\label{tab:Outlook}
\end{sidewaystable}

\subsection{Implications on high-performance computing}
The adoption of this approach translates to the following set-up of the operational analysis and forecast suites at weather centres. Today, the operational, ML-inference based production suites can be run on a few GPUs in seconds-minutes. Due to more sophisticated ML architectures, the focus on ensemble predictions, and an increase in resolution and complexity (e.g., adding ocean, land, snow and ice components) will cause a significant increase in the computing cost. However, ML inference will still be much cheaper when compared to today's physics based models. Re-training of the largest models and foundation models from scratch would only be necessary when scientifically justified, but it could be afforded multiple times per year. For smaller-sized domains/products, re-training is much cheaper and could be updated more frequently, as required by specific products and applications. And it can focus on the frequent training of so-called tail networks that adjust the large and generic foundation models to specific application needs.

However, the generation of training datasets will generate very large cost. As an example, we take the plans for the future ECMWF reanalysis ERA-6 that is funded by the European Commission's Copernicus programme. ERA reanalyses rely on numerical model and data assimilation advances provided by ECMWF's operational system and add previously inaccessible observational datasets and features helping to create a seamless multi-decadal, physically consistent time series of weather \citep{hersbach2020era5}. Such global reanalyses provide boundary conditions for limited-area systems that focus on regions of special interest at higher resolution but usually covering shorter time periods (e.g., in the Arctic: \citep{bromwich2016comparison}). These systems are usually not prepared in real time like operational suites. 

According to first estimates (H. Hersbach, pers. communication), ERA-6 is expected to run its main analysis and medium-range forecast suites as an 11-member ensemble, of which the control analysis occupies at least 12 HPC nodes and the 10 perturbed members probably slightly less. If we assume 100 nodes for the ensemble to be run in multiple parallel streams for the period 1950-2025, it would take 4 streams and a continuous allocation of 400 nodes to complete 40 analysis years in 1.5 years, or 80 years in 3 years. A more ambitious set-up with more parallel streams, higher resolution and with bigger ensembles would clearly challenge present NWP centre HPC capacities as this workload is added on top of the daily operational tasks.

Given the production time and the time needed for the user community's uptake, new versions are only produced every 8 years or so today. This means that model version, spatial resolution, ensemble size etc. are lagging behind operational systems by about 8-10 years at the end of a cycle. Since forecast skill presently increases by a rate of about 1 day of lead time per decade \citep{bauer2015quiet}, creating new reanalysis versions at least every 2-3 years is justified if it wasn't for the substantial computing effort and user uptake. The new scenario should aim to achieve this target.

Faster progress in NWP also means a capacity for running highly realistic, i.e., high-resolution, more complex Earth-system models that presently represent strategic targets in 5-10 years rather than become operational with the next system upgrade. Apart from running research experiments towards this goal, also longer-period trial suites in near-operational settings need to be executed in a timely manner to perform solid performance and stability assessments. If km-scale coupled atmosphere-ocean-land simulations with a throughput of 1 simulated year per day are the target, the requirement for node allocations would be of the order of an entire machine with heterogeneous, very powerful CPU-GPU nodes and high-speed interconnects like Alps at CSCS (see Table~\ref{tab:HPC}). The same applies to high-resolution 40--80-year reanalyses that should be produced within less than a year. This is not affordable now or in the near future by individual centres. This also applies to the resulting data handling challenge \citep{hoefler2023earth}, where flexible memory and storage systems have to cater for both fast-access/high-throughput users and long-term storage.  

Advancing the speed of the creation of these reference datasets will require investments in acceleration by ML inside the numerical models, but also deployment on much larger infrastructures than individual NWP centres can afford. These datasets are quickly becoming one of the most important assets of the NWP community, and their accelerated production requires investments in infrastructure -- and in new ways of collaboration.

\subsection{Implications on collaborations}
\begin{figure}[h]
    \centering
    \includegraphics[scale=0.25]{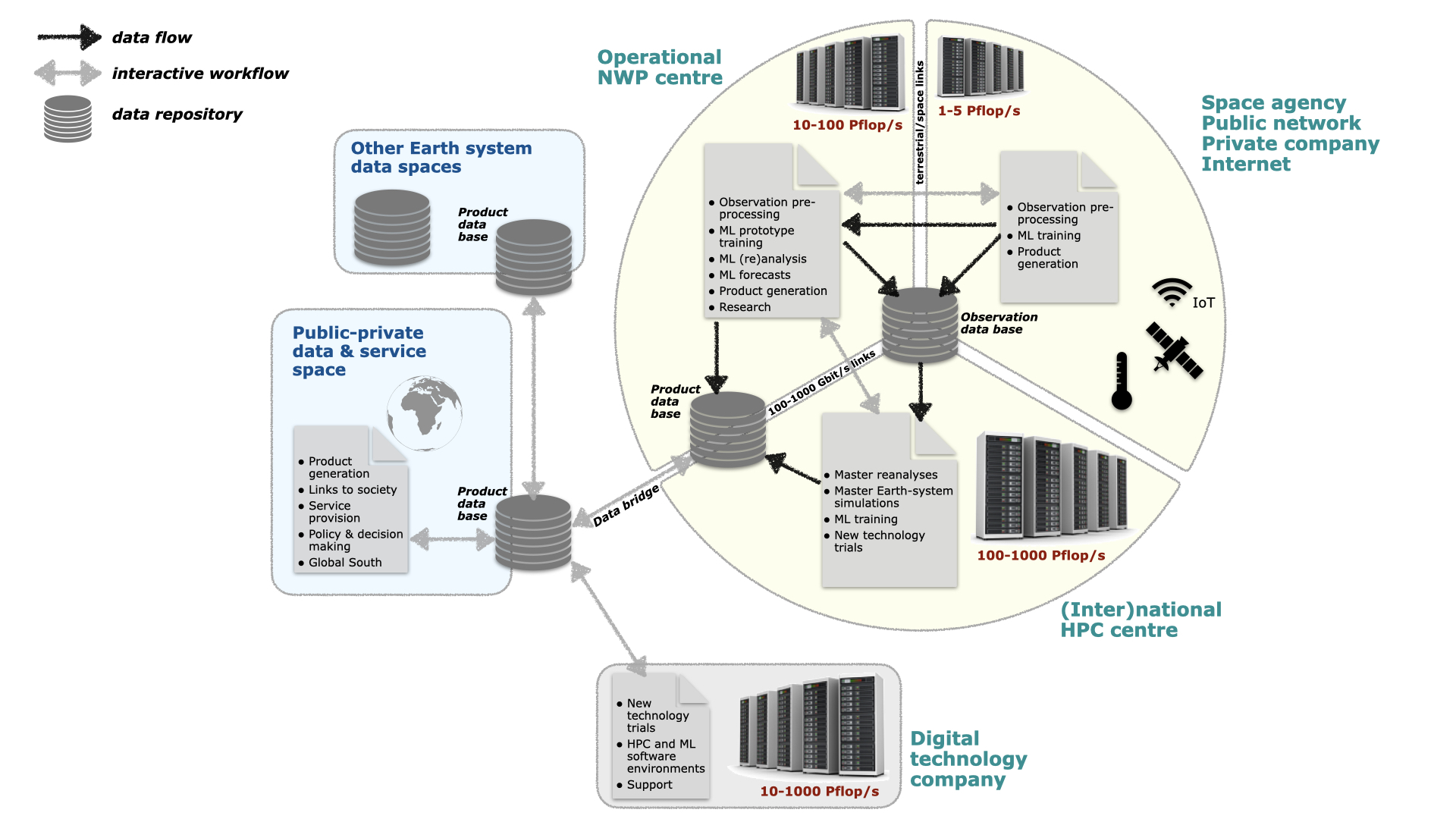}
    \caption{Conceptual model of future work distribution framework between operational NWP centres, data providers and national/international HPC centres, which also interface with digital technology companies, the wider public-private service ecosystem and data spaces. Note that both data providers and HPC centres have further links to other data spaces and public-private services that are not shown here.}
    \label{fig:Collab}
\end{figure}

Figure~\ref{fig:Collab} provides a concept for the interplay between selected operational NWP and HPC centres as well as data providers. This concept deviates from the existing single-entity focus and introduces much more collaboration on computing and data handling to collectively stem the imminent challenges.

The conceptual model would only be effective for those NWP centres, which are in charge of producing the most demanding analysis and forecast datasets. Such centres would be in charge of the co-production of these datasets at extreme scale, but actually outsource the production itself to HPC centres. It would comprise very high-resolution coupled reanalyses and very high-resolution simulations providing the most realistic simulations of weather. This requires the combined expertise of the global and limited-area numerical weather prediction in addition to hydrology, ocean and land surface communities, but eventually extend into Earth-system science as a whole to cover climate space-time scales. Generating these so-called master datasets requires advanced HPC expertise to adapt to the leading technologies, but also a selection and testing process to maximise data quality and minimise computing cost.

The conceptual model aims to minimize data movement. There would be shared fast-access and high-throughput data handling facilities near the HPC centres, and NWP centres but also external users would be able to perform data operations and ML training tasks on these infrastructures without the need to copy data (grey arrows in Figure~\ref{fig:Collab}). The model encourages shared observation and analysis/forecast product databases. To minimize data transfer there would only be a few physical databases placed near the production sites to which observational or model output data stream (black arrows in Figure~\ref{fig:Collab}). Selected subsets of data can be transmitted via high-speed links. 

Analyses and reanalyses only using observations without additional forecasts could be among the ultimate outcomes. Today's operational observational database at ECMWF amounts to approximately 150 GByte/day including 800 million observations. This data has already been quality controlled and reduced to avoid redundancy. Once produced at the NWP centre, they can be easily shared with other data analytics users, even if this volume would grow by, say, an order of magnitude in the next few years. As a host and to democratise usage, the observational database in Figure~\ref{fig:Collab} would become a central part of inter-agency collaboration and between data providers and users to make ML methods more effective.

The data would serve the wider community and spawn new research. Its generation be increasingly delegated to HPC centres having the extreme-scale HPC capacities that NWP centres will ultimately lack. The master datasets will increasingly include ML components and, depending on the ongoing research, be complemented by foundation models. This should limit the HPC needs of NWP centres to about where they are at present and, as more and more cost-effective ML workloads provide the operational output, release sufficient resources for ML training and physics based research at home. Based on these master datasets, specific applications needing additional data and models can be placed at the master sites.

Several components of this conceptual model are already being developed and tested. For example, the option of running NWP suites within flexible software and cluster configuration environments at multiple HPC centres is being prototyped at the Swiss CSCS \citep{alam2023versatile}. Their so-called vcluster (versatile software-defined clusters) allows to allocate specific portions of the same HPC system to workloads that require different (even hardware dependent) software environments and therefore also allow separate change management. This infrastructure-as-code approach introduces much more flexibility for the portability of complex workloads between machines. Since Switzerland is part of the LUMI consortium \citep{LUMICons}, such an approach could extend into northern European countries that can also more easily fulfil the sustainable power and cooling requirements for very large systems. This would allow to run large-scale, generic workloads across different HPC architectures in Europe rather soon.

The Destination Earth data bridges are another example (see Figure~\ref{fig:Collab}). They sit next to the presently largest European HPC infrastructures called LUMI, Leonardo and MareNostrum5 (see also Table~\ref{tab:HPC}). These data bridges come with several PBytes worth of storage both inside and outside the firewall-protected HPC systems that users can access. Data is continuously streamed from the HPC production suites onto these bridges, and users can make use of a vast range of data management and ML tools as part of Destination Earth's so-called digital-twin engine \citep{DestinEDTEngine} without transferring much data. 

Similar open-access infrastructures could be placed near other centres following this template and also deploying similar software and data analytics services. ECMWF is already installing a data hub in Bologna and a high-speed datalink to Alps at CSCS that is similar to such a bridge. This will demonstrate the distributed technical set-up tailored to ML-training for NWP, and prepare for similar designs on EuroHPC machines as part of Destination Earth and the wider NWP community.

All these systems also provide access to federated and generic cloud services, to which post-processing and other production tasks can be delegated, avoiding the need to move datasets in the PByte/day range. The Copernicus Climate Change Service \citep{CDS} can serve as an example how open data is efficiently made available to a very large user base. There are also ideas for scaling this up to a wider federated data handling concept \citep{hoefler2023earth}.

An essential input for the evolution towards our scenario (shown in Table~\ref{tab:Outlook}) and the community co-production effort will be the collection and standardisation of the observational data records from operational and commercial providers as well as data from individual field campaigns. Ideally, this data pool is continually updated, documented and replenished. There is vast experience with this task in the NWP community, and the existing space agency and WMO mechanisms should be sufficient to manage the process. But they may need to adapt faster to the technological progress, user requirements, and commercial interests.

The implication on governance of resources owned by national extreme-scale HPC centres and/or administered by internationally funded allocations (e.g., EuroHPC) needs to be addressed with urgency. Highest priority has the commitment to making resources available under the strict operational constraints of NWP services. The workloads in this concept would go beyond classical one-off research experiments managed by competitive calls national HPC centres are familiar with. We require sustainable allocations of thousands of GPU nodes for weeks to months to achieve the necessary throughput. And, computing resource management has to go hand in hand with data management. In our scenario, the HPC centres need an approach for securing sustainable allocations and an operational workflow and dataflow management system that interfaces with the NWP-centres for these allocations. The cost and governance for this resource could be shared between countries. 

Since the leading HPC centres collaborate with digital technology companies through their contracts and a well established network, these companies can also support the growing NWP workloads with access to new technology, even running selected large workloads and through co-developing software infrastructures. In turn, the commercial sector can benefit from master datasets for training ML methods feeding their business models. Defining quality and ethics standards for NWP data driven models but also large-language models that will help interpret data and user interaction together would be a great achievement \citep{bauer2023deep}. 

\section{Concluding remarks}\label{Concl}
This paper has been motivated by the concern that the operational production set-up of NWP centres, mostly based on individually owning and fully operating HPC and software infrastructures, will soon reach affordability limits. There is a need for faster progress in NWP given the challenges that high-impact weather and climate change impose on society. This need translates into an enhanced community investment in science and the enabling technology. ML offers great opportunities for accelerating service production and enhancing service quality. This paper provides a concept for how to build on the existing community frameworks to set up an ambitious monitoring and forecasting configuration that best exploits the existing digital technologies.

It is likely that the drivers of computing and data handling needs in NWP will soon be the generation of ML training datasets based on km-scale model simulations and comprehensive reanalysis records constrained by all available observations, but also foundation models. These will ultimately require HPC resources that individual NWP centres can not afford. They will even stretch the limits of the largest national HPC centres. 

The preferred use of ML suites in time-critical operational production would enhance efficiency and more flexibility with regards to the ingestion of new science and new products. It does not mean, however, that traditional, first-principles based numerical modeling loses importance as ML merely simplifies production. Also the most advanced physics based simulations reach computing and data limits, but can be managed more easily outside time-critical production and through efforts concerted between NWP centres, a wide range of top-tier HPC centres and digital technology providers. The ultimate target is to run much more ambitious simulations and reanalyses than presently possible, but not at the expense of daily production. 

This demands a new level of coordination and dependencies between centres, and we argue that the outsourcing of the largest computing and data handling demands onto the available HPC infrastructures needs to be co-funded by several countries, go beyond meteorological agencies and happen in co-development with digital technology companies to benefit from the latest technology. 

The link between public entities and commercial companies providing both digital technologies and methodological ML solutions relies on mutual benefits. If future extreme-scale HPC architectures will entirely rely on low-precision (e.g., 16 bit or less) processing to support artificial-intelligence applications, the generation of reference simulations and reanalyses will not be supported on these machines. And, if companies start patenting their algorithms the presently rather open exchange on fundamental ML science may struggle to survive. Weather extremes and climate change are important enough to invest in partnerships that provide for both science development and commercial interests.

Our concept would break the single centre -- single vendor approach that will ultimately limit cost-effectiveness. If the shared NWP-HPC centre concept is applied in various countries at the same time, it will benefit from a wider range of technologies and solutions and can react with more agility to the fast paced technology evolution. There are existing developments in Europe, funded by national programmes and the European Commission, that can be used to prepare such a federated framework. In principle, our collaboration model adds a new stage to why ECMWF has been founded 50 years ago: pool resources to move beyond what individual centres could do, except that pooling in a single place would now become a wider federation. This approach will have implications on digital technology programmes of the European Commission though as these are not built to run operational services.

Also the climate modeling community is looking into concepts that produce more cost-effective decadal and centennial simulations on the largest HPC infrastructures in the world. This may need a NWP-type approach for developing the next generation of models and transferring them into production, but also making the data available more quickly and with more data analytics support \citep{stevens2023earth}. There are many common elements between weather and climate prediction in terms of accelerating science developments and reference dataset generation, and for defining the role ML can play therein \citep{eyring2024pushing}. Some of the solutions in our concept may help both communities, which would be an important efficiency gain. 

\section*{Author contribution}
\PB conceptualised and wrote this paper. 

\section*{Competing interests}
The author declares that there are no competing interests. 


\section*{Acknowledgements}
The author would like to thank Peter Dueben (ECMWF) for his substantial contributions to this work and Torsten Hoefler (ETH Z\"urich) for providing valuable comments and proposals for improvement. He would also like to thank Hans Hersbach (ECMWF) for sharing first estimates of the ERA-6 configuration, acknowledging that these may still change, and thank Martin Janousek (ECMWF) for producing Fig.~\ref{fig:scores}.

\bibliographystyle{plainnat}
\bibliography{JEMS_Final}

\begin{thebibliography}{67}
\providecommand{\natexlab}[1]{#1}
\providecommand{\url}[1]{\texttt{#1}}
\expandafter\ifx\csname urlstyle\endcsname\relax
  \providecommand{\doi}[1]{doi: #1}\else
  \providecommand{\doi}{doi: \begingroup \urlstyle{rm}\Url}\fi

\bibitem[Alam et~al.(2023)Alam, Gila, Klein, Martinasso, and Schulthess]{alam2023versatile}
Sadaf~R Alam, Miguel Gila, Mark Klein, Maxime Martinasso, and Thomas~C Schulthess.
\newblock Versatile software-defined hpc and cloud clusters on alps supercomputer for diverse workflows.
\newblock \emph{The International Journal of High Performance Computing Applications}, 37\penalty0 (3-4):\penalty0 288--305, 2023.

\bibitem[Bauer et~al.(2015)Bauer, Thorpe, and Brunet]{bauer2015quiet}
Peter Bauer, Alan Thorpe, and Gilbert Brunet.
\newblock The quiet revolution of numerical weather prediction.
\newblock \emph{Nature}, 525\penalty0 (7567):\penalty0 47--55, 2015.

\bibitem[Bauer et~al.(2021)Bauer, Dueben, Hoefler, Quintino, Schulthess, and Wedi]{bauer2021digital}
Peter Bauer, Peter~D Dueben, Torsten Hoefler, Tiago Quintino, Thomas~C Schulthess, and Nils~P Wedi.
\newblock The digital revolution of earth-system science.
\newblock \emph{Nature Computational Science}, 1\penalty0 (2):\penalty0 104--113, 2021.

\bibitem[Bauer et~al.(2023)Bauer, Dueben, Chantry, Doblas-Reyes, Hoefler, McGovern, and Stevens]{bauer2023deep}
Peter Bauer, Peter Dueben, Matthew Chantry, Francisco Doblas-Reyes, Torsten Hoefler, Amy McGovern, and Bjorn Stevens.
\newblock Deep learning and a changing economy in weather and climate prediction.
\newblock \emph{Nature Reviews Earth \& Environment}, 4\penalty0 (8):\penalty0 507--509, 2023.

\bibitem[Bi et~al.(2022)Bi, Xie, Zhang, Chen, Gu, and Tian]{pangu}
Kaifeng Bi, Lingxi Xie, Hengheng Zhang, Xin Chen, Xiaotao Gu, and Qi~Tian.
\newblock Pangu-weather: A 3d high-resolution model for fast and accurate global weather forecast.
\newblock \emph{arXiv preprint arXiv:2211.02556}, 2022.

\bibitem[Bodnar et~al.(2024)Bodnar, Bruinsma, Lucic, Stanley, Brandstetter, Garvan, Riechert, Weyn, Dong, Vaughan, et~al.]{aurora}
Cristian Bodnar, Wessel~P Bruinsma, Ana Lucic, Megan Stanley, Johannes Brandstetter, Patrick Garvan, Maik Riechert, Jonathan Weyn, Haiyu Dong, Anna Vaughan, et~al.
\newblock Aurora: A foundation model of the atmosphere.
\newblock \emph{arXiv preprint arXiv:2405.13063}, 2024.

\bibitem[Bonavita and Laloyaux(2020)]{Bonavita}
Massimo Bonavita and Patrick Laloyaux.
\newblock Machine learning for model error inference and correction.
\newblock \emph{Journal of Advances in Modeling Earth Systems}, 12\penalty0 (12):\penalty0 e2020MS002232, 2020.
\newblock \doi{https://doi.org/10.1029/2020MS002232}.
\newblock URL \url{https://agupubs.onlinelibrary.wiley.com/doi/abs/10.1029/2020MS002232}.
\newblock e2020MS002232 10.1029/2020MS002232.

\bibitem[Bouall{\`e}gue et~al.(2024)Bouall{\`e}gue, Weyn, Clare, Dramsch, Dueben, and Chantry]{bouallegue2024}
Zied~Ben Bouall{\`e}gue, Jonathan~A Weyn, Mariana~CA Clare, Jesper Dramsch, Peter Dueben, and Matthew Chantry.
\newblock Improving medium-range ensemble weather forecasts with hierarchical ensemble transformers.
\newblock \emph{Artificial Intelligence for the Earth Systems}, 3\penalty0 (1):\penalty0 e230027, 2024.

\bibitem[Bouallègue et~al.(2024)Bouallègue, Clare, Magnusson, Gascón, Maier-Gerber, Janoušek, Rodwell, Pinault, Dramsch, Lang, Raoult, Rabier, Chevallier, Sandu, Dueben, Chantry, and Pappenberger]{ZiedPangu}
Zied~Ben Bouallègue, Mariana C~A Clare, Linus Magnusson, Estibaliz Gascón, Michael Maier-Gerber, Martin Janoušek, Mark Rodwell, Florian Pinault, Jesper~S Dramsch, Simon T~K Lang, Baudouin Raoult, Florence Rabier, Matthieu Chevallier, Irina Sandu, Peter Dueben, Matthew Chantry, and Florian Pappenberger.
\newblock The rise of data-driven weather forecasting: A first statistical assessment of machine learning-based weather forecasts in an operational-like context.
\newblock \emph{Bulletin of the American Meteorological Society}, 2024.
\newblock \doi{10.1175/BAMS-D-23-0162.1}.
\newblock URL \url{https://journals.ametsoc.org/view/journals/bams/aop/BAMS-D-23-0162.1/BAMS-D-23-0162.1.xml}.

\bibitem[Bromwich et~al.(2016)Bromwich, Wilson, Bai, Moore, and Bauer]{bromwich2016comparison}
David~H Bromwich, Aaron~B Wilson, Le-Sheng Bai, George~WK Moore, and Peter Bauer.
\newblock A comparison of the regional arctic system reanalysis and the global era-interim reanalysis for the arctic.
\newblock \emph{Quarterly Journal of the Royal Meteorological Society}, 142\penalty0 (695):\penalty0 644--658, 2016.

\bibitem[Brunet et~al.(2023)Brunet, Parsons, Ivanov, Lee, Bauer, Bernier, Bouchet, Brown, Busalacchi, Flatter, et~al.]{brunet2023advancing}
Gilbert Brunet, David~B Parsons, Dimitar Ivanov, Boram Lee, Peter Bauer, Natacha~B Bernier, Veronique Bouchet, Andy Brown, Antonio Busalacchi, Georgina~Campbell Flatter, et~al.
\newblock Advancing weather and climate forecasting for our changing world.
\newblock \emph{Bulletin of the American Meteorological Society}, 104\penalty0 (4):\penalty0 E909--E927, 2023.

\bibitem[Carman et~al.(2017)Carman, Clune, Giraldo, Govett, Gross, Kamrathe, Lee, McCarren, Michalakes, Sandgathe, et~al.]{carman2017position}
Jessie Carman, Thomas Clune, Francis Giraldo, M~Govett, Brian Gross, A~Kamrathe, Tsengdar Lee, David McCarren, John Michalakes, Scott Sandgathe, et~al.
\newblock Position paper on high performance computing needs in earth system prediction. national earth system prediction capability.
\newblock Technical report, Technical Report. Retrived from https://doi. org/10.7289/V5862DH3, 2017.

\bibitem[Carrassi et~al.(2018)Carrassi, Bocquet, Bertino, and Evensen]{carrassi2018data}
Alberto Carrassi, Marc Bocquet, Laurent Bertino, and Geir Evensen.
\newblock Data assimilation in the geosciences: An overview of methods, issues, and perspectives.
\newblock \emph{Wiley Interdisciplinary Reviews: Climate Change}, 9\penalty0 (5):\penalty0 e535, 2018.

\bibitem[Cheon and Mun(2023)]{cheon2023climate}
Minjong Cheon and Changbae Mun.
\newblock The climate of innovation: Ai’s growing influence in weather prediction patents and its future prospects.
\newblock \emph{Sustainability}, 15\penalty0 (24):\penalty0 16681, 2023.

\bibitem[ClimateAI(2024)]{ClimateAI}
ClimateAI.
\newblock Climateai has u.s. patent granted for genai-based approach applied to weather forecasting, 2024.
\newblock \url{https://climate.ai/blog/climateai-patent-genai-applied-to-weather-forecasting/#:~:text=This%20newly%20patented%20ClimateAi%20system,biases%20in%20current%20weather%20models.}

\bibitem[Commission(2024)]{DestinE}
European Commission.
\newblock Building a highly accurate digital twin of the earth, 2024.
\newblock \url{https://destination-earth.eu}.

\bibitem[Copernicus(2024)]{CDS}
Copernicus.
\newblock The copernicus climate data store, 2024.
\newblock \url{https://cds.climate.copernicus.eu/#!/home}.

\bibitem[CSC(2023)]{LUMICons}
CSC.
\newblock Lumi consortium, 2023.
\newblock \url{https://www.lumi-supercomputer.eu/lumi-consortium}.

\bibitem[Dahm et~al.(2023)Dahm, Davis, Deconinck, Elbert, George, McGibbon, Wicky, Wu, Kung, Ben-Nun, et~al.]{dahm2023pace}
Johann Dahm, Eddie Davis, Florian Deconinck, Oliver Elbert, Rhea George, Jeremy McGibbon, Tobias Wicky, Elynn Wu, Christopher Kung, Tal Ben-Nun, et~al.
\newblock Pace v0. 2: a python-based performance-portable atmospheric model.
\newblock \emph{Geoscientific Model Development}, 16\penalty0 (9):\penalty0 2719--2736, 2023.

\bibitem[Dongarra et~al.(2003)Dongarra, Luszczek, and Petitet]{dongarra2003linpack}
Jack~J Dongarra, Piotr Luszczek, and Antoine Petitet.
\newblock The linpack benchmark: past, present and future.
\newblock \emph{Concurrency and Computation: practice and experience}, 15\penalty0 (9):\penalty0 803--820, 2003.

\bibitem[Dueben and Bauer(2018)]{dueben2018challenges}
Peter~D Dueben and Peter Bauer.
\newblock Challenges and design choices for global weather and climate models based on machine learning.
\newblock \emph{Geoscientific Model Development}, 11\penalty0 (10):\penalty0 3999--4009, 2018.

\bibitem[ECMWF(2024{\natexlab{a}})]{Cycle49r1}
ECMWF.
\newblock Implementation of ifs cycle 49r1, 2024{\natexlab{a}}.
\newblock \url{https://confluence.ecmwf.int/display/FCST/Implementation+of+IFS+Cycle+49r1#ImplementationofIFSCycle49r1-Keyconfigurationchanges}.

\bibitem[ECMWF(2024{\natexlab{b}})]{Cycle49r1HPC}
ECMWF.
\newblock How to optimise the computing aspects of numerical weather forecasts, 2024{\natexlab{b}}.
\newblock \url{https://www.ecmwf.int/en/about/media-centre/news/2024/how-optimise-computing-aspects-numerical-weather-forecasts}.

\bibitem[ECMWF(2024{\natexlab{c}})]{DestinEDTEngine}
ECMWF.
\newblock The digital twin engine, 2024{\natexlab{c}}.
\newblock \url{https://stories.ecmwf.int/the-digital-twin-engine/}.

\bibitem[ECMWF(2024{\natexlab{d}})]{ECMWFmonitoring}
Reading/bonn/Bologna ECMWF.
\newblock Ecmwf observational data monitoring, 2024{\natexlab{d}}.
\newblock \url{https://www.ecmwf.int/en/forecasts/quality-our-forecasts/monitoring-observing-system#Satellite}.

\bibitem[Eyring et~al.(2024)Eyring, Collins, Gentine, Barnes, Barreiro, Beucler, Bocquet, Bretherton, Christensen, Dagon, et~al.]{eyring2024pushing}
Veronika Eyring, William~D Collins, Pierre Gentine, Elizabeth~A Barnes, Marcelo Barreiro, Tom Beucler, Marc Bocquet, Christopher~S Bretherton, Hannah~M Christensen, Katherine Dagon, et~al.
\newblock Pushing the frontiers in climate modelling and analysis with machine learning.
\newblock \emph{Nature Climate Change}, pages 1--13, 2024.

\bibitem[Frolov et~al.(2024)Frolov, Garrett, Jankov, Kleist, Stewart, and Ten~Hoeve]{frolov2024integration}
Sergey Frolov, Kevin Garrett, Isidora Jankov, Daryl Kleist, Jebb~Q Stewart, and John Ten~Hoeve.
\newblock Integration of emerging data-driven models into the noaa research to operation pipeline for numerical weather prediction.
\newblock \emph{Bulletin of the American Meteorological Society}, 2024.

\bibitem[Gates et~al.(1999)Gates, Boyle, Covey, Dease, Doutriaux, Drach, Fiorino, Gleckler, Hnilo, Marlais, et~al.]{gates1999overview}
W~Lawrence Gates, James~S Boyle, Curt Covey, Clyde~G Dease, Charles~M Doutriaux, Robert~S Drach, Michael Fiorino, Peter~J Gleckler, Justin~J Hnilo, Susan~M Marlais, et~al.
\newblock An overview of the results of the atmospheric model intercomparison project (amip i).
\newblock \emph{Bulletin of the American Meteorological Society}, 80\penalty0 (1):\penalty0 29--56, 1999.

\bibitem[Geer(2021)]{geer2021learning}
Alan~J Geer.
\newblock Learning earth system models from observations: machine learning or data assimilation?
\newblock \emph{Philosophical Transactions of the Royal Society A}, 379\penalty0 (2194):\penalty0 20200089, 2021.

\bibitem[Govett et~al.(2024)Govett, Bah, Bauer, Berod, Bouchet, Corti, Davis, Duan, Graham, Honda, et~al.]{govett2024exascale}
Mark Govett, Bubacar Bah, Peter Bauer, Dominique Berod, Veronique Bouchet, Susanna Corti, Chris Davis, Yihong Duan, Tim Graham, Yuki Honda, et~al.
\newblock Exascale computing and data handling: Challenges and opportunities for weather and climate prediction.
\newblock \emph{Bulletin of the American Meteorological Society}, 2024.

\bibitem[Harris et~al.(2022)Harris, McRae, Chantry, Dueben, and Palmer]{harris2022}
Lucy Harris, Andrew~TT McRae, Matthew Chantry, Peter~D Dueben, and Tim~N Palmer.
\newblock A generative deep learning approach to stochastic downscaling of precipitation forecasts.
\newblock \emph{Journal of Advances in Modeling Earth Systems}, 14\penalty0 (10):\penalty0 e2022MS003120, 2022.

\bibitem[Hersbach et~al.(2020)Hersbach, Bell, Berrisford, Hirahara, Hor{\'a}nyi, Mu{\~n}oz-Sabater, Nicolas, Peubey, Radu, Schepers, et~al.]{hersbach2020era5}
Hans Hersbach, Bill Bell, Paul Berrisford, Shoji Hirahara, Andr{\'a}s Hor{\'a}nyi, Joaqu{\'\i}n Mu{\~n}oz-Sabater, Julien Nicolas, Carole Peubey, Raluca Radu, Dinand Schepers, et~al.
\newblock The era5 global reanalysis.
\newblock \emph{Quarterly Journal of the Royal Meteorological Society}, 146\penalty0 (730):\penalty0 1999--2049, 2020.

\bibitem[Hewitt et~al.(2022)Hewitt, Fox-Kemper, Pearson, Roberts, and Klocke]{hewitt2022small}
Helene Hewitt, Baylor Fox-Kemper, Brodie Pearson, Malcolm Roberts, and Daniel Klocke.
\newblock The small scales of the ocean may hold the key to surprises.
\newblock \emph{Nature Climate Change}, 12\penalty0 (6):\penalty0 496--499, 2022.

\bibitem[Hoefler et~al.(2023)Hoefler, Stevens, Prein, Baehr, Schulthess, Stocker, Taylor, Klocke, Manninen, Forster, et~al.]{hoefler2023earth}
Torsten Hoefler, Bjorn Stevens, Andreas~F Prein, Johanna Baehr, Thomas Schulthess, Thomas~F Stocker, John Taylor, Daniel Klocke, Pekka Manninen, Piers~M Forster, et~al.
\newblock Earth virtualization engines: a technical perspective.
\newblock \emph{Computing in Science \& Engineering}, 25\penalty0 (3):\penalty0 50--59, 2023.

\bibitem[Hohenegger et~al.(2023)Hohenegger, Korn, Linardakis, Redler, Schnur, Adamidis, Bao, Bastin, Behravesh, Bergemann, et~al.]{hohenegger2023icon}
Cathy Hohenegger, Peter Korn, Leonidas Linardakis, Ren{\'e} Redler, Reiner Schnur, Panagiotis Adamidis, Jiawei Bao, Swantje Bastin, Milad Behravesh, Martin Bergemann, et~al.
\newblock Icon-sapphire: simulating the components of the earth system and their interactions at kilometer and subkilometer scales.
\newblock \emph{Geoscientific Model Development}, 16\penalty0 (2):\penalty0 779--811, 2023.

\bibitem[HPCWire(2021)]{MetOfficeMicrosoft}
HPCWire.
\newblock Behind the met office’s procurement of a billion-dollar microsoft system, 2021.
\newblock \url{https://www.hpcwire.com/2021/05/13/behind-the-met-offices-procurement-of-a-billion-dollar-microsoft-system/}.

\bibitem[HPCWire(2023)]{NOAAGDIT}
HPCWire.
\newblock Gdit expands noaa supercomputing capacity for advanced national weather forecasting, 2023.
\newblock \url{https://www.hpcwire.com/off-the-wire/gdit-expands-noaa-supercomputing-capacity-for-advanced-national-weather-forecasting/}.

\bibitem[Huang et~al.(2024)Huang, Gianinazzi, Yu, Dueben, and Hoefler]{huang2024diffda}
Langwen Huang, Lukas Gianinazzi, Yuejiang Yu, Peter~D Dueben, and Torsten Hoefler.
\newblock Diffda: a diffusion model for weather-scale data assimilation.
\newblock \emph{arXiv preprint arXiv:2401.05932}, 2024.

\bibitem[Kochkov et~al.(2023)Kochkov, Yuval, Langmore, Norgaard, Smith, Mooers, Lottes, Rasp, D{\"u}ben, Kl{\"o}wer, et~al.]{neuralgcm}
Dmitrii Kochkov, Janni Yuval, Ian Langmore, Peter Norgaard, Jamie Smith, Griffin Mooers, James Lottes, Stephan Rasp, Peter D{\"u}ben, Milan Kl{\"o}wer, et~al.
\newblock Neural general circulation models.
\newblock \emph{arXiv preprint arXiv:2311.07222}, 2023.

\bibitem[Laloyaux et~al.(2022)Laloyaux, Kurth, Dueben, and Hall]{Laloyaux}
Patrick Laloyaux, Thorsten Kurth, Peter~Dominik Dueben, and David Hall.
\newblock Deep learning to estimate model biases in an operational nwp assimilation system.
\newblock \emph{Journal of Advances in Modeling Earth Systems}, 14\penalty0 (6):\penalty0 e2022MS003016, 2022.
\newblock \doi{https://doi.org/10.1029/2022MS003016}.
\newblock URL \url{https://agupubs.onlinelibrary.wiley.com/doi/abs/10.1029/2022MS003016}.
\newblock e2022MS003016 2022MS003016.

\bibitem[Lam et~al.(2022)Lam, Sanchez-Gonzalez, Willson, Wirnsberger, Fortunato, Alet, Ravuri, Ewalds, Eaton-Rosen, Hu, et~al.]{graphcast}
Remi Lam, Alvaro Sanchez-Gonzalez, Matthew Willson, Peter Wirnsberger, Meire Fortunato, Ferran Alet, Suman Ravuri, Timo Ewalds, Zach Eaton-Rosen, Weihua Hu, et~al.
\newblock Graphcast: Learning skillful medium-range global weather forecasting.
\newblock \emph{arXiv preprint arXiv:2212.12794}, 2022.

\bibitem[Lang et~al.(2024)Lang, Alexe, Chantry, Dramsch, Pinault, Raoult, Clare, Lessig, Maier-Gerber, Magnusson, et~al.]{aifs}
Simon Lang, Mihai Alexe, Matthew Chantry, Jesper Dramsch, Florian Pinault, Baudouin Raoult, Mariana~CA Clare, Christian Lessig, Michael Maier-Gerber, Linus Magnusson, et~al.
\newblock Aifs-ecmwf's data-driven forecasting system.
\newblock \emph{arXiv preprint arXiv:2406.01465}, 2024.

\bibitem[Lawrence et~al.(2018)Lawrence, Rezny, Budich, Bauer, Behrens, Carter, Deconinck, Ford, Maynard, Mullerworth, et~al.]{lawrence2018crossing}
Bryan~N Lawrence, Michael Rezny, Reinhard Budich, Peter Bauer, J{\"o}rg Behrens, Mick Carter, Willem Deconinck, Rupert Ford, Christopher Maynard, Steven Mullerworth, et~al.
\newblock Crossing the chasm: how to develop weather and climate models for next generation computers?
\newblock \emph{Geoscientific Model Development}, 11\penalty0 (5):\penalty0 1799--1821, 2018.

\bibitem[Lessig et~al.(2023)Lessig, Luise, Gong, Langguth, Stadler, and Schultz]{atmorep}
Christian Lessig, Ilaria Luise, Bing Gong, Michael Langguth, Scarlet Stadler, and Martin Schultz.
\newblock Atmorep: A stochastic model of atmosphere dynamics using large scale representation learning.
\newblock \emph{arXiv preprint arXiv:2308.13280}, 2023.

\bibitem[McGovern et~al.(2022)McGovern, Ebert-Uphoff, Gagne, and Bostrom]{mcgovern2022we}
Amy McGovern, Imme Ebert-Uphoff, David~John Gagne, and Ann Bostrom.
\newblock Why we need to focus on developing ethical, responsible, and trustworthy artificial intelligence approaches for environmental science.
\newblock \emph{Environmental Data Science}, 1:\penalty0 e6, 2022.

\bibitem[Michalakes(2020)]{michalakes2020hpc}
John Michalakes.
\newblock Hpc for weather forecasting.
\newblock \emph{Parallel Algorithms in Computational Science and Engineering}, pages 297--323, 2020.

\bibitem[M{\"u}ller et~al.(2019)M{\"u}ller, Deconinck, K{\"u}hnlein, Mengaldo, Lange, Wedi, Bauer, Smolarkiewicz, Diamantakis, Lock, et~al.]{muller2019escape}
Andreas M{\"u}ller, Willem Deconinck, Christian K{\"u}hnlein, Gianmarco Mengaldo, Michael Lange, Nils Wedi, Peter Bauer, Piotr~K Smolarkiewicz, Michail Diamantakis, Sarah-Jane Lock, et~al.
\newblock The escape project: energy-efficient scalable algorithms for weather prediction at exascale.
\newblock \emph{Geoscientific Model Development}, 12\penalty0 (10):\penalty0 4425--4441, 2019.

\bibitem[Nguyen et~al.(2023)Nguyen, Brandstetter, Kapoor, Gupta, and Grover]{climax}
Tung Nguyen, Johannes Brandstetter, Ashish Kapoor, Jayesh~K Gupta, and Aditya Grover.
\newblock Climax: A foundation model for weather and climate.
\newblock \emph{arXiv preprint arXiv:2301.10343}, 2023.

\bibitem[Pacchiardi et~al.(2024)Pacchiardi, Adewoyin, Dueben, and Dutta]{pacchiardi2024}
Lorenzo Pacchiardi, Rilwan~A Adewoyin, Peter Dueben, and Ritabrata Dutta.
\newblock Probabilistic forecasting with generative networks via scoring rule minimization.
\newblock \emph{Journal of Machine Learning Research}, 25\penalty0 (45):\penalty0 1--64, 2024.

\bibitem[Palmer(2020)]{palmer2020vision}
Tim Palmer.
\newblock A vision for numerical weather prediction in 2030.
\newblock \emph{arXiv preprint arXiv:2007.04830}, 2020.

\bibitem[Palmer and Stevens(2019)]{palmer2019scientific}
Tim Palmer and Bjorn Stevens.
\newblock The scientific challenge of understanding and estimating climate change.
\newblock \emph{Proceedings of the National Academy of Sciences}, 116\penalty0 (49):\penalty0 24390--24395, 2019.

\bibitem[Pathak et~al.(2022)Pathak, Subramanian, Harrington, Raja, Chattopadhyay, Mardani, Kurth, Hall, Li, Azizzadenesheli, et~al.]{fourcastnet}
Jaideep Pathak, Shashank Subramanian, Peter Harrington, Sanjeev Raja, Ashesh Chattopadhyay, Morteza Mardani, Thorsten Kurth, David Hall, Zongyi Li, Kamyar Azizzadenesheli, et~al.
\newblock Fourcastnet: A global data-driven high-resolution weather model using adaptive fourier neural operators.
\newblock \emph{arXiv preprint arXiv:2202.11214}, 2022.

\bibitem[Price et~al.(2023)Price, Sanchez-Gonzalez, Alet, Ewalds, El-Kadi, Stott, Mohamed, Battaglia, Lam, and Willson]{gencast}
Ilan Price, Alvaro Sanchez-Gonzalez, Ferran Alet, Timo Ewalds, Andrew El-Kadi, Jacklynn Stott, Shakir Mohamed, Peter Battaglia, Remi Lam, and Matthew Willson.
\newblock Gencast: Diffusion-based ensemble forecasting for medium-range weather.
\newblock \emph{arXiv preprint arXiv:2312.15796}, 2023.

\bibitem[Quintino and Smart(2019)]{NextGenIO}
Tiago Quintino and Simon Smart.
\newblock Running ecmwf workflow on the nextgenio prototype, 2019.
\newblock \url{https://events.ecmwf.int/event/143/contributions/940/attachments/227/412/NEXTGenIO-Quntino.pdf}.

\bibitem[Rackow et~al.(2022)Rackow, Danilov, Goessling, Hellmer, Sein, Semmler, Sidorenko, and Jung]{rackow2022delayed}
Thomas Rackow, Sergey Danilov, Helge~F Goessling, Hartmut~H Hellmer, Dmitry~V Sein, Tido Semmler, Dmitry Sidorenko, and Thomas Jung.
\newblock Delayed antarctic sea-ice decline in high-resolution climate change simulations.
\newblock \emph{Nature communications}, 13\penalty0 (1):\penalty0 637, 2022.

\bibitem[Roberts et~al.(2024)Roberts, Reed, Bao, Barsugli, Camargo, Caron, Chang, Chen, Christensen, Danabasoglu, et~al.]{roberts2024high}
Malcolm~John Roberts, Kevin~A Reed, Qing Bao, Joseph~J Barsugli, Suzana~J Camargo, Louis-Philippe Caron, Ping Chang, Cheng-Ta Chen, Hannah~M Christensen, Gokhan Danabasoglu, et~al.
\newblock High resolution model intercomparison project phase 2 (highresmip2) towards cmip7.
\newblock \emph{EGUsphere}, 2024:\penalty0 1--41, 2024.

\bibitem[Schulthess et~al.(2018)Schulthess, Bauer, Wedi, Fuhrer, Hoefler, and Sch{\"a}r]{schulthess2018reflecting}
Thomas~C Schulthess, Peter Bauer, Nils Wedi, Oliver Fuhrer, Torsten Hoefler, and Christoph Sch{\"a}r.
\newblock Reflecting on the goal and baseline for exascale computing: a roadmap based on weather and climate simulations.
\newblock \emph{Computing in Science \& Engineering}, 21\penalty0 (1):\penalty0 30--41, 2018.

\bibitem[Shalf(2020)]{shalf2020future}
John Shalf.
\newblock The future of computing beyond moore’s law.
\newblock \emph{Philosophical Transactions of the Royal Society A}, 378\penalty0 (2166):\penalty0 20190061, 2020.

\bibitem[Stevens et~al.(2019)Stevens, Satoh, Auger, Biercamp, Bretherton, Chen, D{\"u}ben, Judt, Khairoutdinov, Klocke, et~al.]{stevens2019dyamond}
Bjorn Stevens, Masaki Satoh, Ludovic Auger, Joachim Biercamp, Christopher~S Bretherton, Xi~Chen, Peter D{\"u}ben, Falko Judt, Marat Khairoutdinov, Daniel Klocke, et~al.
\newblock Dyamond: the dynamics of the atmospheric general circulation modeled on non-hydrostatic domains.
\newblock \emph{Progress in Earth and Planetary Science}, 6\penalty0 (1):\penalty0 1--17, 2019.

\bibitem[Stevens et~al.(2023)Stevens, Adami, Ali, Anzt, Aslan, Attinger, B{\"a}ck, Baehr, Bauer, Bernier, et~al.]{stevens2023earth}
Bjorn Stevens, Stefan Adami, Tariq Ali, Hartwig Anzt, Zafer Aslan, Sabine Attinger, Jaana B{\"a}ck, Johanna Baehr, Peter Bauer, Natacha Bernier, et~al.
\newblock Earth virtualization engines (eve).
\newblock \emph{Earth System Science Data Discussions}, 2023:\penalty0 1--14, 2023.

\bibitem[Sun et~al.(2024)Sun, Zhong, Xu, Huang, Li, Feng, Han, Wu, and Qi]{sun2024fuxi}
Xiuyu Sun, Xiaohui Zhong, Xiaoze Xu, Yuanqing Huang, Hao Li, Jie Feng, Wei Han, Libo Wu, and Yuan Qi.
\newblock Fuxi weather: An end-to-end machine learning weather data assimilation and forecasting system.
\newblock \emph{arXiv preprint arXiv:2408.05472}, 2024.

\bibitem[Targett et~al.(2021)Targett, Luk, Lange, and Marsden]{targett2021systematically}
James~Stanley Targett, Wayne Luk, Michael Lange, and Olivier Marsden.
\newblock Systematically migrating an operational microphysics parameterisation to fpga technology.
\newblock In \emph{2021 IEEE 29th Annual International Symposium on Field-Programmable Custom Computing Machines (FCCM)}, pages 69--77. IEEE, 2021.

\bibitem[Vaughan et~al.(2024)Vaughan, Markou, Tebbutt, Requeima, Bruinsma, Andersson, Herzog, Lane, Hosking, and Turner]{aardvark}
Anna Vaughan, Stratis Markou, Will Tebbutt, James Requeima, Wessel~P Bruinsma, Tom~R Andersson, Michael Herzog, Nicholas~D Lane, J~Scott Hosking, and Richard~E Turner.
\newblock Aardvark weather: end-to-end data-driven weather forecasting.
\newblock \emph{arXiv preprint arXiv:2404.00411}, 2024.

\bibitem[Wang et~al.(2024{\natexlab{a}})Wang, Wang, Hu, Wang, Huo, Wang, Wang, Wang, Zhu, Xu, et~al.]{xihe}
Xiang Wang, Renzhi Wang, Ningzi Hu, Pinqiang Wang, Peng Huo, Guihua Wang, Huizan Wang, Sengzhang Wang, Junxing Zhu, Jianbo Xu, et~al.
\newblock Xihe: A data-driven model for global ocean eddy-resolving forecasting.
\newblock \emph{arXiv preprint arXiv:2402.02995}, 2024{\natexlab{a}}.

\bibitem[Wang et~al.(2024{\natexlab{b}})Wang, Tsaris, Liu, Choi, Fan, Zhang, Yin, Ashfaq, Lu, and Balaprakash]{orbit}
Xiao Wang, Aristeidis Tsaris, Siyan Liu, Jong-Youl Choi, Ming Fan, Wei Zhang, Junqi Yin, Moetasim Ashfaq, Dan Lu, and Prasanna Balaprakash.
\newblock Orbit: Oak ridge base foundation model for earth system predictability.
\newblock \emph{arXiv preprint arXiv:2404.14712}, 2024{\natexlab{b}}.

\bibitem[Wedi(2014)]{wedi2014increasing}
Nils~P Wedi.
\newblock Increasing horizontal resolution in numerical weather prediction and climate simulations: illusion or panacea?
\newblock \emph{Philosophical Transactions of the Royal Society A: Mathematical, Physical and Engineering Sciences}, 372\penalty0 (2018):\penalty0 20130289, 2014.

\bibitem[Wedi et~al.(2020)Wedi, Polichtchouk, Dueben, Anantharaj, Bauer, Boussetta, Browne, Deconinck, Gaudin, Hadade, et~al.]{wedi2020baseline}
Nils~P Wedi, Inna Polichtchouk, Peter Dueben, Valentine~G Anantharaj, Peter Bauer, Souhail Boussetta, Philip Browne, Willem Deconinck, Wayne Gaudin, Ioan Hadade, et~al.
\newblock A baseline for global weather and climate simulations at 1 km resolution.
\newblock \emph{Journal of Advances in Modeling Earth Systems}, 12\penalty0 (11):\penalty0 e2020MS002192, 2020.

\end{thebibliography}

\end{document}